\documentclass[a4paper,11pt]{article}
\usepackage{graphicx}
\usepackage[dvips]{color}

\renewcommand{\r}{({\bf r})}

\newcommand{\be}{\begin{equation}}
\newcommand{\ee}{\end{equation}}
\newcommand{\bea}{\begin{eqnarray}}
\newcommand{\eea}{\end{eqnarray}}

\newcommand{\lmt}{\left(\begin{array}{cc}}
\newcommand{\lmf}{\left(\begin{array}{cccc}}
\newcommand{\rma}{\end{array}\right)}
\newcommand{\lvec}{\left(\begin{array}{c}}
\newcommand{\rvec}{\end{array}\right)}
\newcommand{\bi}{\bibitem}

\newcommand{\la}{\langle}
\newcommand{\ra}{\rangle}

\newcommand{\tcb}[1]{\textcolor{blue}{#1}}
\newcommand{\tcr}[1]{\textcolor{red}{#1}}
\newcommand{\tcg}[1]{\textcolor{green}{#1}}

\begin{document}
                                                                   
\title{\bf Density-functional treatment of model Hamiltonians: 
basic concepts and application to the Heisenberg model\footnote{This is 
the (slightly modified) English translation of a chapter, originally written 
in Portuguese, of a proceedings volume on density-functional theory:
{\em 40 anos DFT: Uma perspectiva Brasileira}, eds. K. Capelle, A. J. R. da 
Silva and A. Fazzio.}}

\author{Valter L. L\'{\i}bero and Klaus W. Capelle\\
Departamento de F\'{\i}sica e Inform\'atica\\
Instituto de F\'{\i}sica de S\~ao Carlos\\
Universidade de S\~ao Paulo\\
Caixa Postal 369, 13560-970 S\~ao Carlos, SP, Brazil}

\date{\today}

\maketitle

\thispagestyle{empty}

\begin{abstract}
We describe how density-functional theory, well-known for its many 
uses in {\em ab initio} calculations of electronic structure, can be used 
to study the ground state of inhomogeneous model Hamiltonians. The basic 
ideas and concepts are discussed for the particular case of  the 
Heisenberg model. As representative applications, illustrating scope 
and limitations of the procedure, we calculate the ground-state energy 
of one-, two- and three-dimensional antiferromagnetic Heisenberg models 
in the presence of boundaries and of impurities in the bulk and at the 
surfaces. Correlations are shown to lift degeneracies present in the 
mean-field approximation. Comparison with exact (brute force) diagonalization 
shows that the density-functional results are a significant improvement over 
the mean-field ones, at negligible extra computational cost.
\end{abstract}

\newpage
\tableofcontents
\vspace*{2cm}
\listoftables
\vspace*{2cm}
\listoffigures
\newpage

\section{\label{intro}Introduction}

 Traditional density-functional theory (DFT) is based on the 
Hohenberg-Kohn theorem, according to which all observables of a quantum 
many-body system are determined uniquely by the ground-state charge 
density \cite{hohenbergkohn,kohnrmp}.
These observables are thus functionals of the density, 
and a major task of DFT is the construction of ever better functionals 
for calculating, e.g., ground-state energies in terms of the 
ground-state density. The local-density approximation (LDA) is one such 
functional, expressing the exchange-correlation ($xc$) energy of an 
inhomogeneous many-particle system in terms of its charge density. 
The density itself is normally calculated by means of a self-consistent 
field (SCF) procedure, solving effective single-particle equations 
known as the Kohn-Sham equations \cite{kohnsham}.

  While widely used in electronic-structure calculations, both in 
solid-state physics and in quantum chemistry, the concepts and tools of 
DFT, such as the Hohenberg-Kohn theorem, the Kohn-Sham equations, and 
the LDA and its many improvements, are much more general, and can be 
applied to many quantum systems not described by the {\em ab initio} 
(Coulomb) Hamiltonian underlying applications of DFT in band-structure and 
quantum-chemical calculations.

  In particular, they can also be applied to model Hamiltonians. In a 
model Hamiltonian description of a physical phenomenon one does not 
primarily focus on material-specific details, but on generic features 
of classes of systems. The use of such models has a long tradition in 
many-body physics and statistical mechanics, and the list of common 
models is long and includes famous examples such as the Ising, 
Heisenberg and Hubbard models.

  DFT can be a useful tool in the study of such models in the presence of 
inhomogeneities \cite{hubbardprl,heisenbergprb1,burke}. Below we briefly 
describe the construction and some representative applications of DFT for the
antiferromagnetic Heisenberg model \cite{heisenbergprb1,heisenbergprb2,sanibel}.
To prepare the ground, we present in Sec. \ref{models} a systematic comparison 
between traditional {\em ab initio}  (Coulomb) DFT 
and DFT for model Hamiltonians. Section \ref{heisenberg} introduces the 
Heisenberg model. In Sec.~\ref{homhemo} we describe expressions for the 
ground-state and correlation energy of homogeneous (spatially uniform) 
Heisenberg models. In Sec.~\ref{inhomhemo} we use these expressions, in the 
spirit of an LDA, to construct a local-spin approximation for inhomogeneous
one, two and three-dimensional Heisenberg models. In Sec.~\ref{applications}
these functionals are used to calculate the ground-state energy of 
inhomogeneous chains, ladders and cubes, and compared with mean-field and 
exact diagonalization data. Sec.~\ref{summary} contains a brief summary and
assessment of the approach.

\newpage

\section{\label{models}Comparison between {\em ab initio} DFT and DFT for model 
Hamiltonians}

In Table \ref{table1} we present a schematic comparison between traditional, 
{\em ab initio}, DFT and DFT for two widely studied model Hamiltonians defined 
on lattices, the Hubbard and the Heisenberg model. The basic idea of lattice 
DFT for model Hamiltonians goes back to Gunnarsson and Sch\"onhammer 
\cite{gs}, in the context of the Hubbard model. A viable parametrization for 
the model's $xc$ energy was proposed in \cite{hubbardprl}, while corresponding 
work for the Heisenberg model is based on \cite{heisenbergprb1}. The following 
paragraphs describe Table \ref{table1}, line by line.

Starting point is, in all cases, a Hohenberg-Kohn (HK) theorem, establishing 
that an intensive density-like quantity, coupled to a suitable generating 
field, determines the ground-state wave function of the model under study. 
While in {\em ab initio} DFT this intensive variable is the charge density
$n\r$, in lattice DFT for the Hubbard model it is the site occupation number
$n_i$, and for the Heisenberg model it is the local spin ${\bf S}_i$.

Having proven a Hohenberg-Kohn theorem, the variational principle allows us 
to obtain the ground-state energy and the ground-state density distribution 
by minimizing a density functional. In {\em ab initio} (Coulomb) DFT and 
Hubbard DFT this 
minimization is normally done indirectly, via self-consistent solution of 
Kohn-Sham (KS) equations. This indirect minimization allows one to circumvent 
having to approximate the single-particle kinetic energy as a density 
functional. The Heisenberg model does not have a kinetic-energy term, and 
the minimization is most conveniently performed directly.

The $xc$ energy is also unknown. Starting point for most approximation schemes 
for the $xc$ functional is the local-density approximation, in which one first 
defines an auxiliary spatially uniform many-body problem, which can be solved 
more easily. In {\em ab initio} DFT this auxiliary many-body system is the 
{\em homogeneous} electron liquid, while for lattice DFTs spatial uniformity 
means equivalent sites. The solution of this uniform auxiliary problem may be 
very difficult, but it is still a lot simpler than the solution of the 
{\em inhomogeneous} many-body problem. In {\em ab initio} DFT the best values 
for the $xc$ energy come from Quantum Monte Carlo simulations \cite{ceperley}. 
The one-dimensional Hubbard model, and the one-dimensional spin one-half 
Heisenberg model, have formally exact solutions by means of the Bethe Ansatz. 
For these systems the $xc$ energy can be extracted, in principle, exactly. 
In higher dimensions or for larger spins this can only be done approximately.

Having solved the spatially uniform system, the usual next step is to
parametrize the $xc$ energy in a form comvenient for numerical calculations.
In {\em ab initio} calculations common parametrizations are those of Vosko,
Wilk and Nusair (VWN) \cite{vwn}, Perdew and Zunger (PZ) \cite{pz}, and Perdew
and Wang (PW) \cite{pw}. For the Hubbard model a viable parametrization was
proposed in \cite{hubbardprl}. Alternative approaches are discussed in
\cite{gs,pastor}. For the Heisenberg model several parametrizations were
proposed in \cite{heisenbergprb1}. Some of these are briefly described below.

Once a parametrization of the per-site $xc$ energy of the uniform model is
available, it can be applied site by site in the corresponding inhomogeneous
model. In {\em ab initio} calculations the inhomogeneity arises from the
crystal
lattice in a solid, or the atomic nuclei in a molecule. In lattice models it
arises from inequivalent sites. For the Heisenberg model three examples of
such lattice-DFT calculations for models with inequivalent sites are given
below. These examples complement those presented in \cite{heisenbergprb2}.

\begin{center}
\begin{table}
\caption[Diferent realizations of the concepts of DFT.]{\label{table1} 
Diferent realizations of the concepts of DFT.
See the main text for a description of the various entries. \protect\\}
\begin{tabular}{c|c|c|c}
   & {\bf Coulomb} & {\bf Hubbard} & {\bf Heisenberg} \\      \hline
HK & $ \Psi[n({\bf r})]$ & $\Psi[n_i]$ & $\Psi[{\bf S}_i]$ \\ \hline
KS & SCF & SCF & ----------- \\   \hline
homog. & electron & equivalent&
equivalent \\
system & liquid & sites & sites \\ \hline
$e^{hom}$ & QMC & BA & SW, BA, DMRG \\ \hline
parametri-& VVW, PZ, PW & PRL {\bf 90}, 146402 & PRB {\bf 68}, 024423 \\
zations of & \cite{vwn,pz,pw}& (2003) \cite{hubbardprl}&
(2003) \cite{heisenbergprb1}  \\
$e^{hom}$ (LDA) & & &  \\ \hline
& inhomog. & inhomog. & inhomog. \\
inhomog. & Coulomb & Hubbard & Heisenberg \\
system & systems & model & model \\ \hline
applications &atoms, molecules  &\multicolumn{2}{c}{inequivalent sites,
CDW, SDW, } \\
& solids ... &\multicolumn{2}{c}{boundaries, impurities,
external fields ...}
\end{tabular}
\end{table}
\end{center}


\section{\label{heisenberg}The Heisenberg model}

One of the requirements on a formalism purporting to describe the 
magnetic properties of matter is that it should predict the existence 
of long-range order of ferro or antiferromagnetic nature \cite{stanley,smart}. 
Within DFT, this requirement is met by spin-density-functional theory (SDFT). 
Within model Hamiltonians, the simplest model displaying such order is 
the Heisenberg model.
\be
\hat{H} = \sum_{\la ij\ra} J_{ij} \hat{\bf S}_i \cdot \hat{\bf S}_j \;,
\label{eq:heisen}
\ee
in which $J_{ij}$ is the "exchange interaction" between spins at 
nearest-neighbour sites. The terminology "exchange" here arises from the 
original motivation for writing Eq.~(\ref{eq:heisen}) as an effective model 
accounting for the consequences of the Pauli exclusion principle 
\cite{heisenberg,dirac}. However, if $J_{ij}$ is calculated from many-body 
theory or fitted to experiment, it generally also contains 
contributions from correlation, i.e., many-body effects beyond the 
single-determinant (Hartree-Fock like) approximation.

  Depending on the sign of $J_{ij}$ and its dependence on the site 
labels $i$ and $j$, the model (\ref{eq:heisen}) can have ferromagnetic, 
antiferromagnetic or noncollinear spin configurations \cite{smart}. 
Below we focus on the antiferromagnetic case, which is relevant for a large 
class of magnetic insulators, among which the undoped state of cuprate 
superconductors may be one of the most important \cite{manousakis}. 
However, the applications of model (\ref{eq:heisen}) transcend the boundaries 
of physics, as illustrated by chemical applications to the magnetic states of 
conjugated hydrocarbons and organometallic compounds [19-24]. 

  Although almost 80 years have passed since the model (\ref{eq:heisen}) 
was first proposed, a complete solution is known only for the limiting cases 
of few sites and, by means of the Bethe Ansatz (BA), for spin $S=1/2$ in one 
dimension \cite{bethe,hulthen}. Complete numerical diagonalization is possible 
for up to a few dozens of sites. The density-matrix renormalization group
(DMRG) and Quantum Monte Carlo (QMC) techniques can be applied 
also to larger systems, but at considerable computational cost.

  Moreover, all these techniques, BA, DMRG, QMC, etc. encounter significant 
difficulties in spatially nonuniform systems, with broken translational 
symmetry, such as in the presence of impurities, boundaries or externally 
applied nonuniform fields. A modification of the homogeneous Heisenberg model
describing this type of inhomogeneity is
\be
\hat{H} = \sum_{ij} J_{ij} \hat{\bf S}_i \cdot \hat{\bf S}_j
+ \sum_i \hat{\bf S}_i {\bf B}_i \;,
\label{eq:inhomheisen}
\ee
where ${\bf B}_i$ is an external or internal magnetic field, which can vary 
from one site to the next, and $J_{ij}$ extends the spin-spin coupling
to other than nearest neighbours.

In these situations DFT provides a 
computationally viable approach to the ground-state energy and density, 
and their dependence on the system parameters. However, as explained in 
the preceding section, already the simplest practical approximation to 
DFT, the LDA, requires as an input the solution of the uniform system. 
For this reason we turn, in the next section, to a brief discussion of 
the ground state and correlation energy of the homogeneous Heisenberg 
model. 
       
\section{\label{homhemo}Ground-state energy of the homogeneous Heisenberg model}

The simplest approximation to Hamiltonian (\ref{eq:heisen}) is the mean-field
(MF) one, in which the vector operators $\hat{\bf S}$ are substituted
by classical (commuting) vectors ${\bf S}$. In this approximation the 
per-spin ground-state energy of a homogeneous linear, quadratic or cubic
lattice comprising $N$ sites with spin $S$, in $d$ dimensions, is
\begin{equation}
e_0^{MF}(S) \equiv \frac{E_0^{MF}(S)}{N}= -Jd\;S^2 \;.
\label{eq:cm} \end{equation}
An improved estimate of this energy can be obtained from spin-wave (SW) theory 
\cite{anderson}, according to which
\begin{eqnarray}
e_0^{SW}(S) = - Jd\;S^2 + Jd^{-1/5} (\frac{2}{\pi}-1)S 
= e_0^{MF}(S) + Jd^{-1/5} (\frac{2}{\pi}-1)S\;.
\end{eqnarray}
(Here we have used the conjecture of Ref.~\cite{heisenbergprb1} to write
the last term as a closed function of $d$.) As an example, for $d=1$ and 
$S=1/2$, spin-wave theory predicts $e_0^{SW}=-0.431690 J$, which is about 
$2.6 \%$ off the exact Bethe-Ansatz result $e_0^{BA}=(1/4-\ln 2)J$. 

The correlation energy $e_c(S)$ is defined as the difference
$e_0(S)-e_0^{MF}(S)$, where $e_0(S)$ is the exact ground-state energy.
Within SW theory, we have thus
\begin{equation}
e_c^{SW}(S) = e_0^{SW}(S) - e_0^{MF}(S) = Jd^ {-1/5}(\frac{2}{\pi}-1)S \;.
\end{equation}
A more precise expression for $e_c(S)$ in $d=1$ has recently been proposed 
\cite{heisenbergprb1,lou} on the basis of previously obtained 
density-matrix renormalization group data \cite{lou},\footnote{The fit
originally proposed in Ref.~\cite{lou} did not include the cubic terms in
$1/S$ and did not recover the Bethe-Ansatz exact result at $S=1/2$. The
cubic terms were added in Ref.~\cite{heisenbergprb1} in order to extend the 
fit to $S=1/2$.}
\begin{eqnarray}
\frac{e_c^{DMRG}(S)}{J} = \frac{e_c^{SW}(S)}{J} - 0.03262 - \frac{0.0030}{S}
 + \frac{0.0015}{S^3} \nonumber \\ -
\left( 0.338 - \frac{0.28}{S} + \frac{0.035}{S^3} \right)
e^{-\pi S} \cos(2\pi S) \;.
\label{eq:ecdmrg}\end{eqnarray}
For $S=1/2$ this expression predicts $e_0^{DMRG}=-0.446253 J$, which deviates
only by $0.7\%$ from the exact result. 

The utility of the MF, SW and DMRG expressions for $e_0(S)$ is, {\it a priori},
rather limited: MF theory is not reliable due to its neglect of correlation,
and the SW and DMRG expressions are only applicable to spatially uniform 
systems, in which $S$ is a constant. (The DMRG expression is also limited to
$d=1$.) Restriction to spatial homogeneity is a serious limitation, since 
spatial inhomogeneities are ubiquitous in real systems. Examples are
effects of external magnetic fields, magneto-crystalline anisotropy, 
boundaries, impurities, defects etc. Such situations are hard to deal
with by traditional methods because translational symmetry is broken.
To overcome this difficulty we employ density-functional theory 
\cite{kohnrmp,dftbook}, within which expressions for the uniform system 
find applications as input for LDA-type approximations for inhomogeneous
systems.

\section{\label{inhomhemo}Ground-state energy of the inhomogeneous Heisenberg 
model: spin-distribution functionals and local-spin approximation}

The Hohenberg-Kohn theorem for the Heisenberg model, proved in 
Ref.~\cite{heisenbergprb1}, shows that the expectation value of any
observable $\hat{O}$ of that model is a functional of the ground-state 
spin distribution ${\bf S}_i$, given, in principle, as $O[{\bf S}_i] = 
\la \Psi[{\bf S}_i] | \hat{O} | \Psi[{\bf S}_i] \ra$, where 
${\bf S}_i=\langle \Psi|\hat{\bf S}_i|\Psi \rangle$ is a classical spin 
vector and not an operator. This functional is universal with respect to 
external fields ${\bf B}_i$, but not with respect to changes in the spin-spin 
interaction $J_{ij}$.

For the Heisenberg model the ground-state energy and its minimizing spin
distribution are obtained most conveniently by direct minimization of the
functional
\begin{equation}
E_0[{\bf S}_i] \equiv E^{MF}_0[{\bf S}_i] + E_c[{\bf S}_i] \;.
\end{equation}
Here the mean-field term is a simple functional of ${\bf S}_i$,
\begin{equation}
E_0^{MF}[{\bf S}_i] = J \sum_{\la ij \ra} {\bf S}_i \cdot {\bf S}_j \;,
\label{cm}
\end{equation}
which for homogeneous antiferromagnetic systems on linear, square and cubic 
lattices reduces to Eq.~(\ref{eq:cm}). To approximate
$E_c[{\bf S}_i]$ we employ the LDA concept, applied directly to the 
spin vectors. The resulting local-spin approximation (LSA) consists in
substituting, locally, the variable $S$ in the expressions for the homogeneous 
system, by the variable $|{\bf S}_i|$, according to
\begin{equation}
E_c[{\bf S}_i] \approx E_c^{LSA}[{\bf S}_i] = 
\sum_i \frac{E_c(S)}{N}|_{S\to |{\bf S}_i|} =:
\sum_i e_c(S)|_{S\to |{\bf S}_i|} \;.
\end{equation} 
Within the SW approximation for $E_c$ we thus have
\begin{equation}
E_c^{LSA-SW}[{\bf S}_i] = J d^{-1/5} \; (\frac{2}{\pi}-1) 
\sum_i |{\bf S}_i| \;,
\end{equation}
so that the ground-state energy functional in the LSA$^{SW}$ becomes
\begin{equation}
E_0^{LSA-SW}[{\bf S}_i] = J\sum_{\la ij \ra} {\bf S}_i \cdot {\bf S}_j
+ J \; d^{-1/5} (\frac{2}{\pi}-1) \sum_i |{\bf S}_i| \;.
\label{eq:eosw} \end{equation}
Similarly, the $LSA^{DMRG}$ is
\be
E_0^{LSA-DMRG}[{\bf S}_i] = J\sum_{\la ij \ra} {\bf S}_i \cdot {\bf S}_j
+ \sum_{\la ij \ra} e_c^{DMRG}(S)|_{S\to |{\bf S}_i|} \;,
\ee
where $e_c^{DMRG}(S)$ is defined in Eq.~(\ref{eq:ecdmrg}).

\section{Applications: Finite-size quantum spin chains, ladders and cubes}
\label{applications}

\begin{figure}
\includegraphics[height=100mm,width=115mm,angle=0]{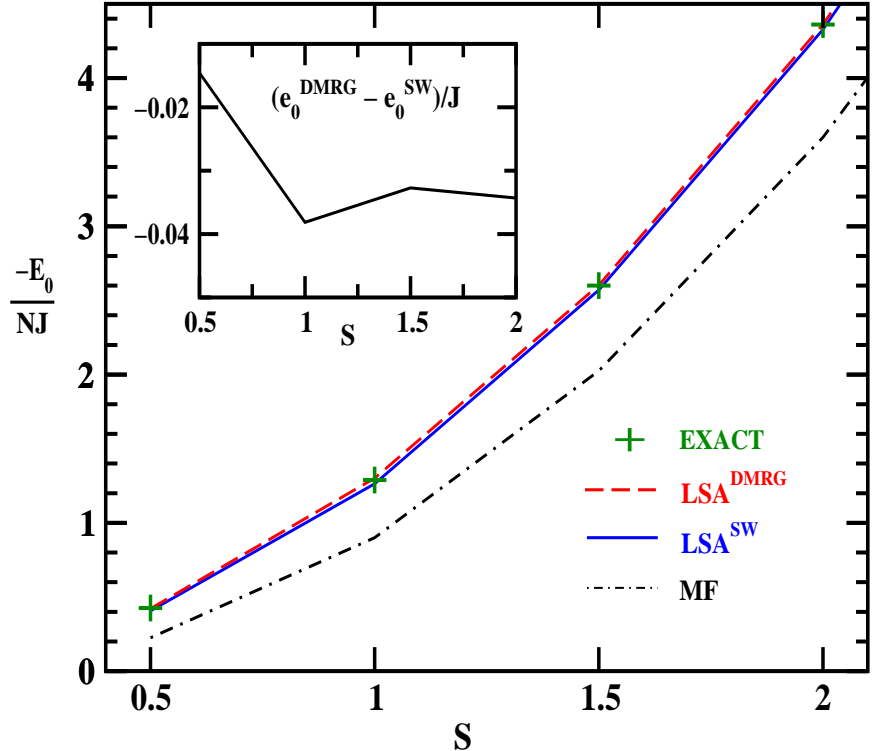}
\caption[Ground-state energy of an open 10-site chain of spin $1/2$ atoms.]
{\label{fig1} Per-site ground-state energy of an open
10-site chain of spin $1/2$ atoms, obtained exactly, in the mean-field
approximation, and with LSA. The inset illustrates the difference
between $LSA^{SW}$ and $LSA^{DMRG}$.}
\end{figure}

As an application of these approximations we now consider 10 spin $S$
sites on a chain with open boundary conditions. As a consequence of the 
boundaries this is already an inhomogeneous system, in which not all sites 
are equivalent. In Fig.~\ref{fig1} we compare values for the per-site 
ground-state energy obtained from the mean-field approximation, with values 
obtained from the LSA$^{SW}$ and LSA$^{DMRG}$ functionals, for some 
values of the spin $S$. We also include exact values obtained from numerical
diagonalization of Hamiltonian (\ref{eq:heisen}). The comparison of the
three approximation schemes with the exact data illustrates the significant
improvement obtained by means of the LSA, as compared to the MF approximation. 
It also illustrates that the LSA$^{SW}$ and LSA$^{DMRG}$ are very similar on 
this scale. In the inset we plot the difference $e_0^{DMRG}-e_0^{SW}$ versus 
$S$, to illustrate the scale of the difference between both formulations of the 
LSA. For larger systems than $L=10$, and/or larger spins than $S=2$, full
numerical diagonalization becomes computationally very expensive. On the other 
hand, cheap mean-field calculations are not reliable. The LSA 
concept is thus a useful tool for estimating energies of systems that are
too large for exact calculations, at much better accuracy than obtained from 
mean-field calculations.

\begin{figure}
\includegraphics[height=100mm,width=125mm,angle=0]{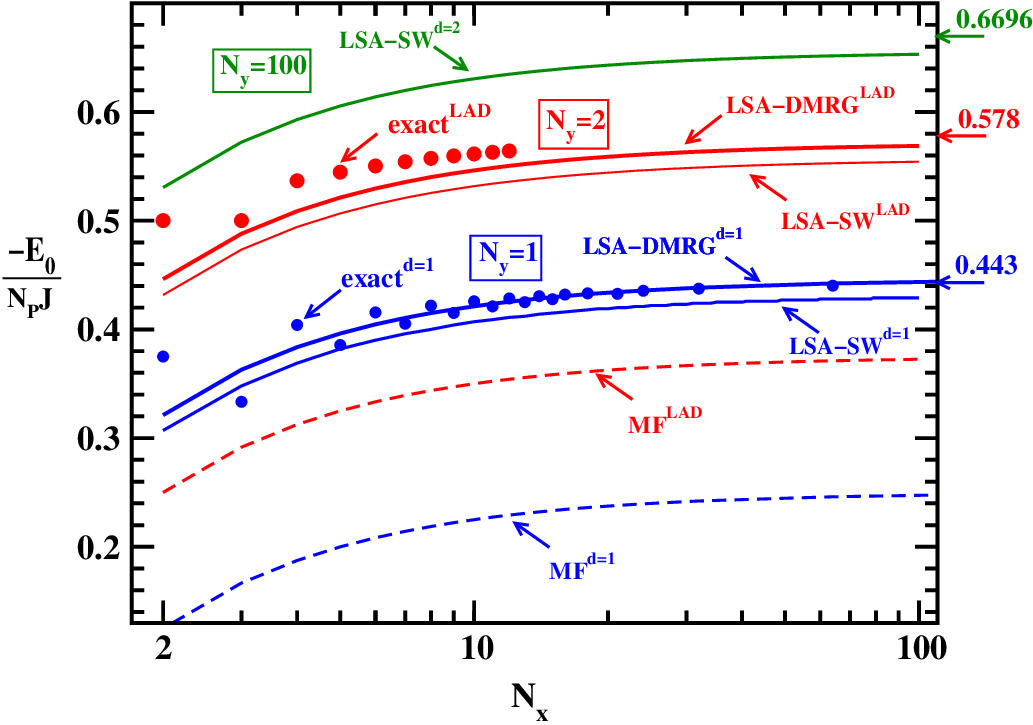}
\caption[Ground-state energy of an $N_x \times N_y$
lattice of $S=1/2$ spins.]
{\label{fig2} Per-site ground-state energy of an $N_y \times N_x$
lattice of $S=1/2$ spins, with open boundary conditions. $N_p=N_y \times N_x$
is the total number of sites, ranging from $2$ to $10^4$. Curves and data
pertaining to a one-dimensional chain of length $N_x$ are in blue,
those pertaining to two-legged ladders of size $2\times N_x$ in red, 
and the curve for multi-legged ladders of size $100\times N_x$ is in green.
See main text for more detailed explanation of the different data sets and 
curves.}
\end{figure}

The main utility of the LSA concept, however, lies in the ease with which
it is extended to higher dimensions or to more inhomogeneous systems. 
As a second illustrative application of the LSA we thus present, in 
Fig.~\ref{fig2}, results for systems representative of the crossover from
one to two dimensions, such as spin ladders. Figure~\ref{fig2} contains two
sets of exact data. The first, labelled exact$^{d=1}$, is the per-site
energy of a linear open chain with $N_x$ sites. For this same system we also
show results from the mean-field approximation, Eq.~(\ref{cm}) (labelled 
MF$^{d=1}$), and the LSA$^{DMRG}$ and LSA$^{SW}$ approaches (labelled 
LSA-DMRG$^{d=1}$ and LSA-SW$^{d=1}$, respectively).
The arrow labelled $0.443$ indicates the exact value in the thermodynamic
limit $N_x\to \infty$, obtained from the Bethe-Ansatz. Comparing these
data we see that the MF approximation fails badly, while LSA$^{DMRG}$ provides
an excellent approximation to the exact data for not too small $N$.

\begin{table}
\caption[Ground-state energy $-E_0/NJ$ in the mean-field approximation, of
a $4 \times 4 \times 4$ cube with 63 spin $1/2$ sites and one impurity spin
$S_I$.]
{\label{table2} Ground-state energy $-E_0/NJ$ in the mean-field approximation,
of a $4 \times 4 \times 4$ cube with 63 spin $1/2$ sites and one spin $S_I$
site located on a corner, an edge, a face or in the bulk.\protect\\}
\begin{tabular}{c|c|c|c|c|c|c}
           & 1/2    & 1      & 3/2    & 2      & 5/2   & 3      \\
\hline
corner& 0.5625 & 0.5742 & \tcb{0.5859} & 0.5977 & \tcg{0.6094} & \tcr{0.6211}\\
edge  & 0.5625 & 0.5781 & 0.5938 & \tcg{0.6094} & 0.6250 & 0.6406 \\
face  & 0.5625 & 0.5820 & 0.6016 & \tcr{0.6211} & 0.6494 & 0.6602 \\
bulk  & 0.5625 & \tcb{0.5859} & \tcg{0.6094} & 0.6328 & 0.6562 & 0.6797
\end{tabular}
\end{table}

\begin{table}
\caption[Ground-state energy $-E_0/NJ$ in the LSA$^{SW}$, of a
$4\times 4\times 4$ cube with 63 spin $1/2$ sites and one imurity spin $S_I$.]
{\label{table3} Ground-state energy $-E_0/NJ$ in the LSA$^{SW}$ of the same
system described in Table~\ref{table2} and Fig.~\ref{fig3}.
Inclusion of correlation lifts the degeneracies predicted by the mean-field
approximation.\\}
\begin{tabular}{c|c|c|c|c|c|c}
           & 1/2    & 1      & 3/2    & 2      & 5/2   & 3      \\
\hline
corner& 0.7080 & 0.7220 & \tcb{0.7360} & 0.7500 & \tcg{0.7640} & \tcr{0.7780}\\
edge  & 0.7080 & 0.7259 & 0.7438 & \tcg{0.7617} & 0.7796 & 0.7975 \\
face  & 0.7080 & 0.7298 & 0.7516 & \tcr{0.7734} & 0.7952 & 0.8170 \\
bulk  & 0.7080 & \tcb{0.7337} & \tcg{0.7594} & 0.7851 & 0.8108 & 0.8366
\end{tabular}
\end{table}

The second set of exact data in Fig.~\ref{fig2}, labelled exact$^{LAD}$, 
represents the energy of spin ladders of size $2 \times N_x$, as a function 
of $N_x$. Again, we compare the exact data with results from the mean field
(MF$^{LAD}$) and both local-spin (LSA-DMRG$^{LAD}$ and LSA-SW$^{d=1}$) 
approximations. The arrow at $0.578$ is the value extrapolated for 
$N_x\to \infty$ from Quantum Monte Carlo data \cite{barnes}. Although for the 
ladders LSA$^{DMRG}$ and LSA$^{SW}$ are not as good as for chains (which is 
easy to understand, considering the origin of their correlation energy in a 
one-dimensional system), it is still a significant improvement on the 
mean-field approximation, which for the two-legged ladder predicts energy 
values that fall closer to the exact ones for the chain than to those for
the ladder itself. In numbers, at $N_x=12$ the MF approximation differs by 
$37 \%$ from the exact value, while the LSA$^{DMRG}$ value is off by $2.4 \%$.
Still in Fig. \ref{fig2}, the curve labelled LSA-SW$^{d=2}$ represents
values for more and more two-dimensional systems, of size $N_x \times 100$,
obtained by combining the two-dimensional mean-field approximation
with the two-dimensional SW approximation as an input for LSA.
The value obtained from LSA$^{SW}$ at $N_x=100$ deviates by only $3.2 \%$ 
from the asymptotic result $0.6696 J$, estimated in Ref.~\cite{liang}.

\begin{figure}
\includegraphics[height=80mm,width=60mm,angle=0]{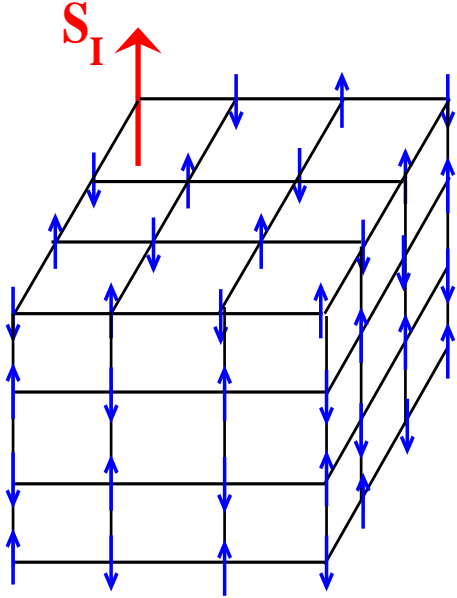}
\caption[$4\times4\times4$ cube of spin $1/2$ with a
substitutional impurity.]
{\label{fig3} $4\times4\times4$ cube of spin $1/2$ with a
substitutional impurity at the corner (illustrated), edge, face or in the
bulk. The mean-field ground-state energies resulting from the different
impurity positions show certain degeneracies that are lifted by inclusion
of correlations. (See data in Table \ref{table2} and \ref{table3} and
discussion in main text.)}
\end{figure}

As our third example we consider a three-dimensional cubic arrangement of
spins with $4\times 4 \times 4 = 64$ spin $1/2$ sites, one of which is
replaced by a substitutional impurity of spin $S_I$. This impurity can sit
in the bulk of the cube (8 equivalent positions), on one of its faces (24
equivalent positions), on an edge (24 equivalent positions), or on the corner
sites (8 equivalent positions). Fig.~\ref{fig3} illustrates this system, and
Tables \ref{table2} and \ref{table3} contain the corresponding mean-field and
LSA$^{SW}$ ground-state energies for several values of $S_I$. We note that the
mean-field approximation predicts certain degeneracies that are lifted on
including correlation via the LSA: in the MF approximation an $S_I=2$ impurity
on a face of the cube and an $S_I=3$ on a corner site are degenerate (both
having energy $e_0^{MF} = -0.6211J$), as are $S_I=1$ impurity in the bulk
and an $S_I=3/2$ one at a corner ($e_0^{MF}=-0.5859J$). A triple degeneracy
occurs between an $S_I=5/2$ impurity at a corner, an $S_I=2$ one on an edge
site, and a $S_I=3/2$ one in the bulk. ($e_0^{MF}=-0.6094J$). All of these
degeneracies are lifted by correlations, which shows that they are artifacts
of the mean-field approximation. This type of information could be useful for
the design of self-assembled magnetic nanostructures, in which the magnetic
atoms will tend to spontaneously occupy the energetically favoured sites.
Other applications of the LSA$^{SW}$ and LSA$^{DMRG}$ functionals can be 
found in Refs.~\cite{heisenbergprb1,heisenbergprb2,sanibel}.

\section{\label{summary} Summary}

The above examples show that DFT within the local-spin approximation 
can be a useful tool for calculating ground-state energies of spatially
inhomogeneous Heisenberg models. The computational effort is much less than 
for a QMC or DMRG calculation, allowing applications to systems in two
or three dimensions, and with reduced translational symmetry. 

Exact or nearly exact benchmark values can only be obtained for small numbers 
of sites and simple geometries, or in the thermodynamic limit.
LSA calculations, on the other hand, are not
more expensive, computationally, than the mean-field ones, and much more
accurate, but this advantage is partially offset by the fact that, unlike an
exact, BA, or QMC calculation, LSA does not yield the many-body wave function.
LSA is also less accurate than DMRG, but more easily applicable to systems in
two and three dimensions.

Methodologically, DFT for the Heisenberg model differs in three aspects 
from Coulomb ({\em ab initio}) DFT \cite{kohnrmp} and from DFT for the
Hubbard model \cite{hubbardprl}: 
(i) LSA proceeds exclusively in terms of the spin density and makes no use of 
the charge density (differently from the local-spin-density approximation of
Coulomb DFT, which uses charge and spin densities).
(ii) Numerical results are obtained by directly minimizing an energy 
functional, and not via self-consistent solution of effective single-particle
equations. In this sense Heisenberg-model DFT in the LSA represents a 
realization of orbital-free-DFT \cite{carter}.
(iii) For the Heisenberg model the dependence of the correlation 
functional on dimensionality is known to a good approximation
\cite{heisenbergprb1}. This dependence is unknown in other formulations
of DFT.

This last observation also represents a different way in which DFT for model
Hamiltonians can be useful: While the above mostly focused on how DFT can
be usefully employed to obtain information on model Hamiltonians, it is
also possible to regard these models as theoretical laboratories in 
which concepts and tools of DFT can be investigated in a simpler and
better controlled environment than in an {\em ab initio} calculation.

{\em Acknowledgements:} This work was supported by FAPESP and CNPq. We 
thank F.~C.~Alcaraz and L.~N.~Oliveira for useful discussions on the Bethe 
Ansatz and the Heisenberg model.


\end{document}